\documentclass[a4paper,11pt]{article}

\usepackage{jheppub} % for details on the use of the package, please
\usepackage[T1]{fontenc} % if needed
\usepackage{braket}
\usepackage{xcolor}
\usepackage{comment}
\newcommand{\XLQ}[1]{\textcolor{red}{#1}}

\title{\boldmath The Coupled SYK model at Finite Temperature}

\author[a,b]{Xiao-Liang Qi}
\author[c,d]{and Pengfei Zhang.}
\affiliation[a]{Stanford Institute for Theoretical Physics, Stanford University}
\affiliation[b]{Department of Physics, Stanford University}
\affiliation[c]{Walter Burke Institute for Theoretical Physics, California Institute of Technology}
\affiliation[d]{Institute for Advanced Study, Tsinghua University}
\abstract{Sachdev-Ye-Kitaev (SYK) model, which describes $N$ randomly interacting Majorana fermions in 0+1 dimension, is found to be an solvable UV-complete toy model for holographic duality in nearly AdS$_2$ dilaton gravity. Ref. \cite{Xiaoliang} proposed a modified model by coupling two identical SYK models, which at  low-energy limit is dual to a global AdS$_2$ geometry. This geometry is an ``eternal wormhole" because the two boundaries are causally connected. Increasing the temperature drives a Hawking-Page like transition from the eternal wormhole geometry to two disconnected black holes with coupled matter field. To gain more understanding of the coupled SYK model, in this work, we study the finite temperature spectral function of this system by numerical solving the Schwinger-Dyson equation in real-time. We find in the low-temperature phase the system is well described by weakly interacting fermions with renormalized single-particle gap, while in the high temperature phase the system is strongly interacting and the single-particle peaks merge. We also study the $q$ dependence of the spectral function.}

\begin{document} 
\maketitle
\flushbottom

\section{Introduction}

In general relativity, traversable wormholes are known to be forbidden by the average null energy condition (ANEC) \cite{ANEC1,ANEC2}. In a recent work, Gao, Jafferis and Wall \cite{Gao} proposed that a traversable wormhole can be induced by turning on a coupling between the two boundaries of the (anti-de Sitter) eternal black hole geometry. The coupling is nonlocal from the bulk point of view, which makes it possible to violate the ANEC. From the boundary point of view (in the sense of holographic duality), the eternal black hole geometry corresponds to the thermal field double state \cite{TFD1,TFD2}, which describes two entangled systems, each has a thermal reduced density matrix. The traversable wormhole describes the fact that scrambled quantum information in one system can be restored from the other system by making use of the coupling and the pre-existing entanglement between them. This is similar to quantum teleportation \cite{susskind2018teleportation,Yoshida,Norman1,Norman2,AdS2,gao2019traversable,brown2019quantum,bak2018bulk,van2018quantum} and is also related to the Hayden-Preskill protocol of recovering information from the black hole \cite{HP}.

%When we throw a few qubits into a black hole, or a generic quantum many-body system, as time evolves the information would eventually get scrambled over the full Hilbert space and no initial data could be extracted unless we measure more than half of the system \cite{Page1,Page2}. However, in an early work \cite{HP}, Hayden and Preskill realize that, in principle, one could extract few-bits quantum states scrambled over the black hole by manipulating a small part of the total system (which represents the Hawking radiation), together with an entangled partner of the black hole. 

%Later, a simple realization for this telepotation process is proposed by Gao, Jafferis and Wall \cite{Gao}\footnote{There is another realization by Yoshida and Kitaev which does not require any specific dynamics of the system \cite{Yoshida}, and has been realized in recent experiments \cite{Norman1,Norman2}.}. They find a coupling (for a finite time window) between two boundaries of an eternal BTZ black hole with definite sign could generate negative energy in bulk. This would make the wormhole traversable, which is usually forbidden in general relativity when the averaged null energy condition is not violated \cite{ANEC1,ANEC2}. In this set-up, the role of the entanglement partner of the left system is naturally played by the right system. And as in the original scenario, we only need to couple some degree of freedom of the left system to its counterpart in the right system. 

In the case of two-dimensional gravity, the dual theory of traversable wormhole has been studied in the Sachdev-Ye-Kitaev (SYK) model \cite{kitaev2015simple,kitaev2018soft,bulk,Kitaev2,Comments,bulk2,bulk3}. The SYK model describes $N$-Majorana fermions with random $q/2$-body coupling, and the low energy physics in the large-$N$ limit agrees with Jackiw-Teitelboim gravity \cite{jackiw1985lower,teitelboim1983gravitation} coupled with matter fields. By considering a thermal field double state of the SYK model, and turning on an instant coupling between the two sides, the traversable wormhole physics can be realized\cite{AdS2,gao2019traversable}. In these models, the traversable wormhole remains open for a finite time, which depends on the strength and time of the coupling term. In contrast, Ref. \cite{Xiaoliang} proposed an ``eternal traversable wormhole", by considering two SYK models with a bilinear coupling, and studying the ground state of the coupled model. By studying the low temperature limit and the large-$q$ limit of the SYK model, Ref. \cite{Xiaoliang} shows that the ground state of the coupled model is approximately the same as the thermal field double state of the (uncoupled) SYK model. Having an eternal traversable wormhole means that a particle from one boundary can reach the other boundary and will oscillate back and forth between the two boundaries. The corresponding bulk geometry is a global AdS$_2$ geometry, with two boundaries that are invariant under the global AdS$_2$ time translation. At finite temperature, the coupled SYK model has a first order phase transition. The low temperature phase is the traversable worm hole, while the high temperature phase can be interpreted as two geometrically disconnected black holes (but with coupled matter fields). A generalization of eternal traversable wormhole in four dimensions was discussed in Ref. \cite{four}. 

In this paper, we study the coupled SYK model beyond the low temperature limit. Although the low energy effective theory approach can predict the first order phase transition \cite{Xiaoliang}, it cannot describe the behavior of matter fields at finite temperature. We expect that particles cannot travel freely at finite temperature, since they scatter with each other. Therefore even in the low temperature phase, the eternal oscillation between two boundaries is probably an approximation to the actual dynamics. Using the Schwinger-Keldysh formalism \cite{Kamenev}, we derive the self-consistent equation for the spectral function matrix of the coupled SYK model. By numerically solving the self-consistent equation, we study the coupling and temperature dependence of the single-particle spectral, focusing on the excitation energy and lifetime. We find that SYK models with small $q$ and large $q$ have quite different finite temperature behavior. For small $q$, the low-temperature phase is well-captured by weakly interacting dilute fermion gas with a temperature-dependent life time, while the high temperature phase has a spectral function without quasiparticle peaks. In the low temperature phase, the excitation energy is consistent with the prediction from low energy effective theory results in Ref. \cite{Xiaoliang}. For large $q$, the spectral function show quasi-particle peaks in both high-temperature and low-temperature phases. This is consistent with the large-$q$ solution which suggests that the two-point function in these two phases are qualitatively the same. However, the quasi-particle energy and life time change discontinuously across the phase transition.

The remainder of this paper is organized as follows. In section \ref{sec:review}, we give a brief review of the SYK model and coupled SYK model. In section \ref{sec:keldysh}, we use the Keldysh approach to derive the self-consistent equation for spectral function and analyze the lifetime by perturbation. Then we discuss the numerical results for $q=4$ and $q=20$ separately in section \ref{sec:q4} and \ref{sec:q20}. Finally, further discussions are presented in Sec. \ref{sec:conclusion}.

\section{Review of the coupled SYK model}\label{sec:review}

The SYK model\cite{Ye,Kitaev2} describes $N$ Majorana fermions $\chi_i$ with random $q/2$ body interactions. (Complex fermion version of this model has also been studied.) The Hamiltonian is written as:
\begin{align}
    H_{\textrm{SYK}}[\chi_i] = ( i ) ^ { q / 2 } \sum _ { 1 \leq j _ { 1 } \leq j _ { 2 } \cdots \leq j _ { q } } J _ { j _ { 1 } j _ { 2 } \cdots j _ { q } } \chi _ { j _ { 1 } } \chi _ { j _ { 2 } } \cdots \chi _ { j _ { q } }.
\end{align}
We take the convention that $\{\chi_i,\chi_j\}=\delta_{ij}$. $J _ { j _ { 1 } j _ { 2 } \cdots j _ { q } } $ with different $j_i$ labels are independent random Gaussian variables (up to anti-symmetrization) with zero mean and the following variants:
\begin{align}
\left\langle J _ { j _ { 1 } \cdots j _ { q } } ^ { 2 } \right\rangle = \frac { 2 ^ { q - 1 } \mathcal { J } ^ { 2 } ( q - 1 ) ! } { q N ^ { q - 1 } }.
\end{align}
For $\beta \mathcal{J}\rightarrow \infty $, the system has emergent conformal symmetry. At low temperature, the low energy effect theory of this model is described by Schwarzian dynamics \cite{Comments}, which is consistent with a holographic dual theory of dilaton gravity. The AdS$_2$ metric can be chosen as
\begin{align}
ds^2=\frac{-dt^2+d\sigma^2}{\sin^2\sigma}.
\end{align}
with $\sigma\in(0,\pi)$. The dilaton field solution breaks the $SL(2,R)$ isometry of AdS$_2$ to a boost symmetry. The boundary is set by a constant dilaton value, which corresponds to the curves
\begin{align}
    \frac{\cos t}{\sin\sigma}=\frac1\epsilon
\end{align}
The two boundaries have finite distance with each other, but remain space-like separated, which represent two systems that are entangled but not coupled.

% In imaginary-time, at the saddle-point, the Green's function is then written as:
% \begin{align}
% G(u)=\left<\mathcal{T}_u\chi_i(u)\chi_i(0)\right>=\frac{c_\Delta \text{sgn}(u)}{(\frac{\beta \mathcal{J}}{\pi}\sin\frac{\pi u}{\beta} )^{2\Delta}},\ \ \ \ \ \ c_{\Delta}=\frac{1}{2}\left((1-2\Delta)\frac{\tan \pi\Delta}{\pi\Delta}\right)
% \end{align}
% where $\Delta=1/q$. The low-energy fluctuations near this saddle-point is found to be described by reparameterization modes $\tau(u)$:
% \begin{align}\label{sykeffective}
% S = - \frac { N\alpha _ { S } } { \mathcal { J } } \int d u  \left\{ \tan \frac { \pi \tau( u ) } { \beta } , u \right\}
% \end{align}
% Here $\{f(u),u\}$ is the Schwarzian derivation defined as $\{f(u),u\}=\frac{f'''}{f'}-\frac{3}{2}\left(\frac{f''}{f'}\right)^2$. 
On comparison, we can consider the same AdS$_2$ geometry with the boundary at 
\begin{align}
    \sin\sigma=\epsilon
\end{align}
which breaks the $SL(2,R)$ symmetry to time translation along $t$ direction. This is called global AdS$_2$ geometry, which is an ``eternal traversible wormhole" because the two boundaries are causally connected. 
%An eternal wormhole is instead described by a global AdS$_2$ space-time. The metric of this space-time is written as:
%\begin{align}
%ds^2=\frac{-dt^2+d\sigma^2}{\sin^2\sigma}.
%\end{align}
%There are two boundaries lie at $\sigma=0$ and $\sigma=\pi$ that could communicate with each other by sending particles. 

It is not possible to realize such a traversible wormhole geometry without violating the average null energy condition\cite{ANEC1,ANEC2}. As is Shown by the work of Gao-Jafferis-Wall\cite{Gao}, it is possible to violate the ANEC and create a traversible wormhole with a finite lifetime by introducing a coupling between the two boundaries. The corresponding physics in SYK model has been discussed in Ref \cite{AdS2,gao2019traversable}.

%To find a SYK-like model that is dual to this geometry in the low-energy limit, naturally we also need two copies $\chi_{i}^L$ and $\chi_{i}^R$ of the original SYK model, each of them would lies on one of the boundaries. 

Based on these developments, Ref. \cite{Xiaoliang} proposed that the eternal traversible wormhole geometry is dual to the ground state of the following coupled SYK Hamiltonian: 

%However, if we assume the matter fields in the bulk is local and obeys the average null energy condition, then the wormhole solution is not possible \cite{ANEC1,ANEC2}. We can break the energy condition by adding coupling terms between the two boundaries:
%\begin{equation}
%    S_{i n t}=g \sum_{i=1}^{N} \int d u O_{L}^{i}(u) O_{R}^{i}(u),
%\end{equation}
%On the SYK side, this motivate use to add direct coupling between two copies $\chi_{i}^L$ and $\chi_{i}^R$.

%With all these ingredients, it is reasonable to introduce the coupled SYK model proposed in \cite{Xiaoliang} with Hamiltonian:

\begin{equation}\label{coupledH}
\begin{aligned}
    H & = H_{\textrm{SYK}}\left[\chi_i^L\right] + (-1)^\frac{q}{2}H_{\textrm{SYK}}\left[\chi_i^R\right] + H_{\textrm{int}} , \,\,\,\, \ \ \  H_{\textrm{int}} = i \mu \sum_{i} \chi_{i}^L \chi_{i}^R.
\end{aligned}
\end{equation}

Without losing generality, we take $\mu>0$. In the zero-temperature limit, for small interaction strength $\mu\ll \mathcal{J}$, Ref.\cite{Xiaoliang} assumed that the low-energy physics is still governed by two reparametrization modes $t_l(u)$ and $t_r(u)$ \cite{Xiaoliang}. In real-time, the effective action is the sum of the Schwarzian term and the coupling term:
\begin{equation}\label{sykeffective}
S = N \int d u \left\{ - \frac { \alpha _ { S } } { \mathcal { J } } \left( \left\{ \tan \frac { t _ { l } ( u ) } { 2 } , u \right\} + \left\{ \tan \frac { t _ { r } ( u ) } { 2 } , u \right\} \right) + \mu \frac { c _ { \Delta } } { ( 2 \mathcal { J } ) ^ { 2 \Delta } } \left[ \frac { t _ { l } ^ { \prime } ( u ) t _ { r } ^ { \prime } ( u ) } { \cos ^ { 2 } \frac { t _ { l } ( u ) - t _ { r } ( u ) } { 2 } } \right] ^ { \Delta } \right\},
\end{equation}
Here $\{f(u),u\}$ is the Schwarzian derivation defined as $\{f(u),u\}=\frac{f'''}{f'}-\frac{3}{2}\left(\frac{f''}{f'}\right)^2$. The saddle point $t_l(u)=t_r(u)=t' u$ gives the Green's function of fermions:
\begin{align} 
&\left<\chi_i^L(u)\chi_i^L(0)\right>=\left<\chi_i^R(u)\chi_i^R(0)\right>=c_\Delta e^{-i\pi\Delta}\left[\frac{t'}{2\mathcal{J}\sin \frac{t'(u-i\epsilon)}{2}}\right]^{2\Delta}, \label{GLLT0}\\
&\left<\chi_i^L(u)\chi_i^R(0)\right>=-\left<\chi_i^R(u)\chi_i^L(0)\right>=ic_\Delta \left[\frac{t'}{2\mathcal{J}\cos \frac{t'u}{2}}\right]^{2\Delta},\label{GLRT0}
\end{align}
with $$c_{\Delta}=\frac{1}{2}\left((1-2\Delta)\frac{\tan \pi\Delta}{\pi\Delta}\right)^\Delta,\ \ \ \ \ \ \ \left(\frac{t'}{\mathcal{J}}\right)^{2(1-\Delta)}=\frac{\mu\Delta}{2\mathcal{J}\alpha_S}\frac{2c_\Delta}{2^{2\Delta}}.$$ Since the real time Green's function is oscillating, we expect the system to be gapped. By expanding Eq. \eqref{GLLT0} in Fourier series, we find the following discrete energy spectrum of fermions:
\begin{align}
E_n^{(m)}=t'(\Delta+n).\label{matter}
\end{align}
which is the same as a bulk field in AdS$2$ space-time. The higher frequency modes have smaller spectral weight. There are also gravitational sector of the spectrum, described by the fluctuation of $t_l(u)$ and $t_r(u)$, which can be calculated to be:
\begin{align}
E_n^{(g)}=t'\sqrt{2(1-\Delta))}\left(n+\frac{1}{2}\right).\label{gravity}
\end{align}
 Comparing Eq. \eqref{matter} and Eq. \eqref{gravity} for $\Delta=1/q$, we find that the gap of the system is determined by the matter field $E_{\text{g}}=t'\Delta$. 

Upon increasing temperature of the coupled system, the coupling term between two sides becomes less important (in the renormalization group sense), and the system would ultimately goes back to two weakly coupled black holes. A first-order transition at $T_c$ is found to separate these two phases by either large $q$ expansion or numerical calculation for finite $q$, which is an analogy of Hawking-Page transition in higher dimension. In the large-$q$ expansion, a third phase which is unstable in canonical ensemble is also found.

\begin{figure}[t]
 	\center
 	\includegraphics[width=0.65\columnwidth]{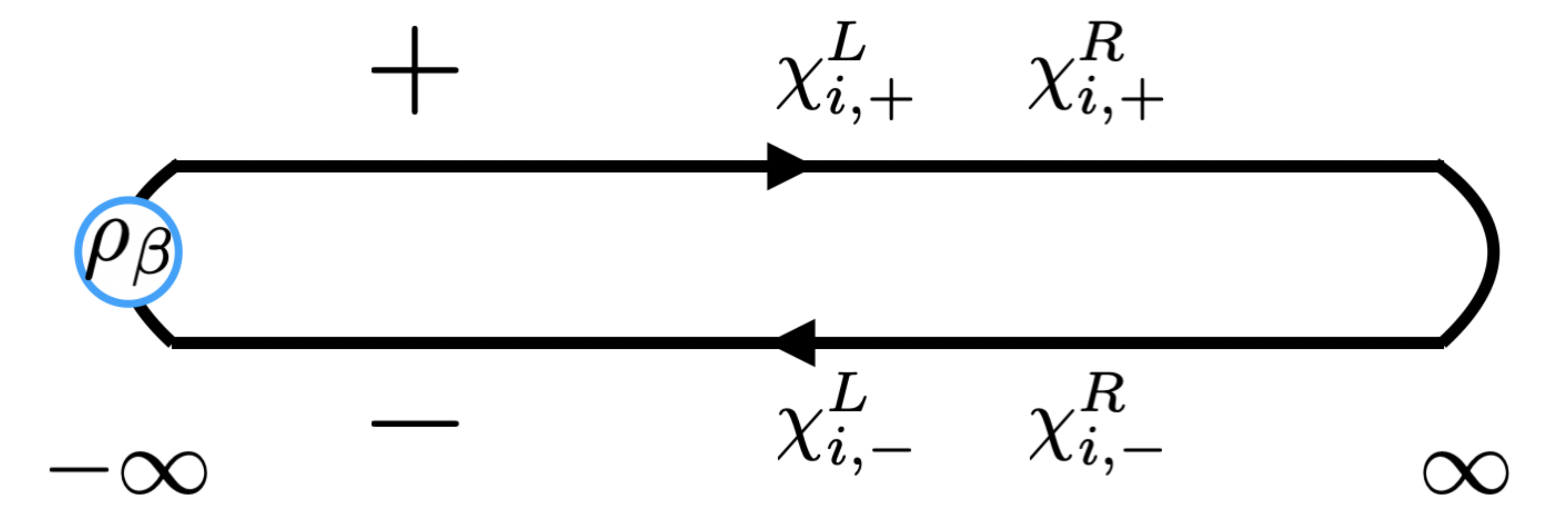}
 	\caption{A sketch for the Keldysh contour with a forward and backward evolution. There are two copies of fields $\chi^L_{i,\pm}$ and $\chi^R_{i,\pm}$.} \label{contour}
 \end{figure}

\section{Self-consistent equation for the spectral function}\label{sec:keldysh}

In this work we focus on the finite temperature behavior of the spectral function of the coupled SYK model, mainly around the transition temperature $T_c$. To derive the self-consistent equation for the spectral function, we directly consider the Keldysh Contour in real-time \cite{Kamenev} with a forward evolution contour and a backward evolution contour as shown in Figure \ref{contour}. Consequently we have two copies of field $\chi^L_{i,\pm}$ and $\chi^R_{i,\pm}$. The generating function is defined as:

\begin{align}
\mathcal{Z}&=\int d JP(J)\int \mathcal{D}\chi^L_+\mathcal{D}\chi^R_+\mathcal{D}\chi^L_-\mathcal{D}\chi^R_-\exp\left(i\int dt L\right),\\
L&=\sum_i\frac{1}{2}\chi^a_{i,\eta}\left((\hat{G}^0)^{-1}\right)^{ab}_{\eta\xi}\chi^a_{i,\xi}-H[\chi^L_{i,+},\chi^R_{i,+}]+H[\chi^L_{i,-},\chi^R_{i,-}]. \label{L}
\end{align}
Here $\eta,\xi=\pm$ and $a,b=L$ or $R$. $\hat{G}^0$ is the bare Green's function for $\mathcal{J}=\mu=0$, defined as:
\begin{align}
\hat{G}_{\eta\xi}^{ab}(t)=-i\left<\chi_{i,\eta}^a(t)\chi_{i,\xi}^b(0)\right>=\begin{pmatrix}
G^{T,ab}(t)&G^{<,ab}(t)\\
G^{>,ab}(t)&G^{\tilde{T},ab}(t)
\end{pmatrix}_{\eta\xi}.
\end{align}
Due to unitarity, all information about two-point function is contained in $G^>$ and $G^<$. Moreover, for Majorana fermions, we have the relation $G^>(t)=(G^<(t))^*$. For simplicity, we assume that $q=4n$ with $n$ an integer. The expression for the self-energy $\hat{\Sigma}=(\hat{G}_{\mathcal{J}=0})^{-1}-\hat{G}^{-1}$ is simple in this basis: 
\begin{align}
\hat\Sigma_{\eta\xi}^{ab}(t)&\equiv\begin{pmatrix}
\Sigma_\chi^{T,ab}(t)&-\Sigma_\chi^{<,ab}(t)\\
-\Sigma_\chi^{>,ab}(t)&\Sigma_\chi^{\tilde{T},ab}(t)
\end{pmatrix}_{\eta\xi}=-\frac{\mathcal{J}^2}{q}\eta \xi \left[2G_{\eta\xi}^{ab}(t)\right]^{q-1}
\end{align}

We are interested in the spectral function $\hat{\rho}(\omega)$ of the system, which is now a matrix in $L/R$ space. It can be determined from the traditional retarded Green's function $G_R^{ab}(t)=-i\theta(t)\left<\{\chi_i^a(t),\chi_i^b(0)\}\right>$ by:
\begin{align}
\hat{\rho}(\omega)=-\frac{1}{2\pi i}(G_R-(G_R)^\dagger) \label{rhodef}
\end{align}
The relation between $\hat{G}$ in the $\pm$ basis and $G_R$ can be determined by standard Keldysh rotation, which gives:
\begin{align}
G_{R}^{ab}(t)&=\theta(t)(G^{>,ab}(t)-G^{<,ab}(t)).\label{GR}
\end{align}
Similarly, the self-energy 
\begin{align}
\Sigma_R=(G_{R,\mathcal{J}=0})^{-1}-G_{R}^{-1}=(\omega+i\epsilon)I+\mu \sigma_y-G_{R}^{-1} \label{self}
\end{align}
is also given as:
\begin{align}
\Sigma_{R}^{ab}(t)&=\theta(t)(\Sigma^{>,ab}(t)-\Sigma^{<,ab}(t))=-\theta(t)\frac{\mathcal{J}^2}{q}\left[(2G^{>,ab}(t))^{q-1}-(2G^{<,ab}(t))^{q-1}\right].\label{SR}
\end{align}
In thermal equilibrium, all two point correlators are just functional of $\hat{\rho}(\omega)$. We have:
\begin{align}
G^{>,ab}(\omega)&=-i\hat{\rho}^{ab}(\omega)n_F(-\omega). \label{G>}
\end{align}
We could then numerically solve Eq. \eqref{rhodef}, Eq. \eqref{self}, Eq. \eqref{SR} and Eq. \eqref{G>} iteratively. Before presenting the numerical results, let's consider what should happen in the low-temperature limit. At the zero-temperature limit Eq. \eqref{GLLT0} and Eq. \eqref{GLRT0}, the spectral function of the system should contain $\delta$-function peaks with zero width, which corresponds to well-defined quasiparticles with infinite lifetime. However, at finite temperature, there should be scattering between quasi-particles and the life-time should become finite. 

To estimate the quasiparticle lifetime, we focus on the mode near $\omega=E_g$ and approximate
%A finite life-time for quasi-particle should come from the self-energy term. We focus on the mode near $\omega=E_g$, and approximate without the scattering process
\begin{align}
G^{(0)}_R(\omega)^{-1}=\begin{pmatrix}
\omega+i\Gamma&-iE_g\\
iE_g&\omega+i\Gamma
\end{pmatrix},
\end{align}
with $\Gamma$ the inverse quasiparticle lifetime. Now we would like to compute $\Gamma$ approximately using the self-consistent equation.

According to Eq. \eqref{rhodef}, this Green's function corresponds to the spectral function 
\begin{align}
\hat{\rho}^{(0)}(\omega)=\frac{1}{2}\begin{pmatrix}
\delta_\Gamma(\omega-E_g)+\delta_\Gamma(\omega+E_g)&i\delta_\Gamma(\omega-E_g)-i\delta_\Gamma(\omega+E_g)\\
-i\delta_\Gamma(\omega-E_g)+i\delta_\Gamma(\omega+E_g)&\delta_\Gamma(\omega-E_g)+\delta_\Gamma(\omega+E_g)
\end{pmatrix}
\end{align}
with $\delta_\Gamma(x)=-\frac1{\pi}{\rm Im}\frac1{\omega+i\Gamma}=\frac1\pi\frac{\Gamma}{\omega^2+\Gamma^2}$.

In the limit $\Gamma\ll E_g,~E_g\beta\gg 1$, the self-energy in Eq. (\ref{SR}) is approximately
\begin{align}
\Sigma^{ab}_R(t)&\propto \mathcal{J}^2\left(n_F(E_g)\right)^{(\frac{q}{2}-1)} \left(n_F(-E_g)\right)^{\frac{q}{2}}\exp(-i E_g t-(q-1)\Gamma t)\nonumber\\ &\sim \mathcal{J}^2\exp\left(-\beta E_g(\frac{q}{2}-1)-i E_g t-(q-1)\Gamma t\right),\\
\Sigma^{ab}_R(\omega)&\sim\frac{\mathcal{J}^2}{\omega-E_g+i(q-1)\Gamma}\exp\left(-\beta E_g(\frac{q}{2}-1)\right).
\end{align}
The imaginary part of this self-energy at $\omega=E_g$ should be $\Gamma$, which means
\begin{align}
\Gamma\sim\text{Im}\Sigma^{ab}_R(\omega)\propto \mathcal{J}^2\exp\left(-\beta E_g(\frac{q}{2}-1)\right)/\Gamma.\nonumber\\
\Gamma\sim \mathcal{J}\exp\left(-\beta E_g\frac{(q/2-1)}{2}\right). \label{Gamma}
\end{align}
Here we have neglected some overall $q$-dependent factors in the inverse life time, and focused on its temperature dependence. This result can also be derived using a semi-classical Boltzmann equation \cite{Kinetics}. Physically, the inverse lifetime is determined by density of other quasiparticles around and their interaction $\mathcal{J}$.

In the next two sections, we will analyze the numerical solution for small $q$ and large $q$, respectively, and compare it with the estimation here.

%To get finite life-time $1/\Gamma\gg1/E_g$, we should focus on $\Sigma_R^{ab}(E_g)$. The contribution comes from the term in \eqref{SR} with frequency $E_g$, which describes an on-shell scattering process with $\frac{q}{2}-1$ incoming excitations and $\frac{q}{2}$ outgoing excitation. In the limit that $E_g\beta\gg 1$, we have:
%\begin{align}
%\Sigma^{ab}_R(t)&\propto \mathcal{J}^2\left(n_F(E_g)\right)^{(\frac{q}{2}-1)} \left(n_F(-E_g)\right)^{\frac{q}{2}}\exp(-i E_g t)\sim \mathcal{J}^2\exp\left(-\beta E_g(\frac{q}{2}-1)-i E_g t\right),\\
%\Sigma^{ab}_R(\omega)&\sim\frac{\mathcal{J}^2}{\omega-E_g+i\epsilon}\exp\left(-\beta E_g(\frac{q}{2}-1)\right).
%\end{align}
%The imaginary part of $\Sigma^{aa}_R(E_g)$ should give a contribution to $\Gamma$. However, in this approximation we find:

\begin{figure}[t]
 	\center
 	\includegraphics[width=1\columnwidth]{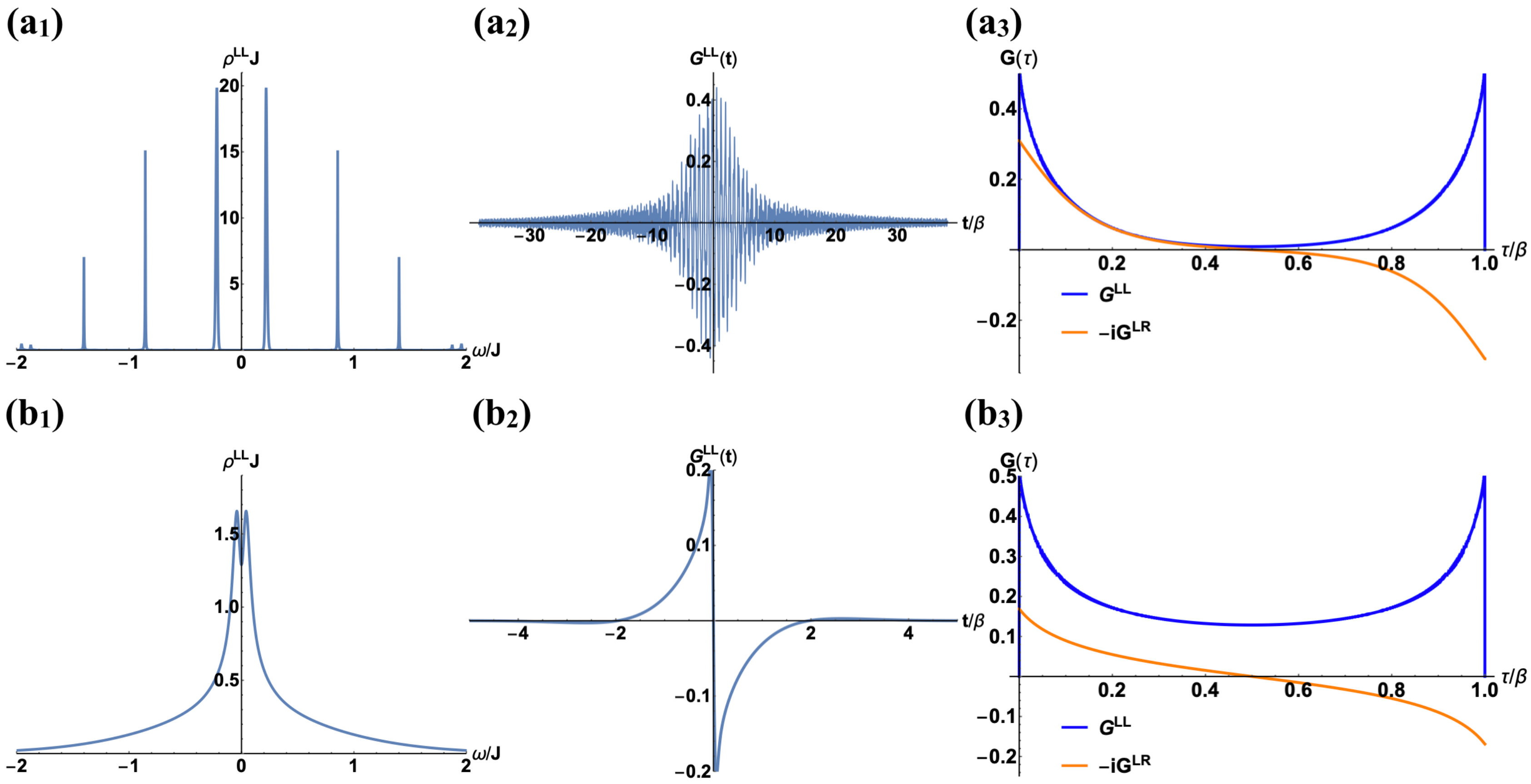}
 	\caption{The spectral function and the Green's function for $q=4$ and $\mu=0.1 \mathcal{J}$. (a$_1$-a$_3$). The spectral function $\rho^{LL}(\omega)$, the real-time Green's function $\text{Re}\ G^{>,LL}(t)$ and the imaginary-time Green's function $G^{ab}(\tau)$ for $\beta \mathcal{J}=40$. The system is in the wormhole phase. (b$_1$-b$_3$). The numerical results of the same quantities for $\beta \mathcal{J}=30$. The system is in the two back holes phase.} \label{fig1}
 \end{figure}

%If the life-time of the quasi-particle becomes finite, we instead expect:
%\begin{align}
%\Gamma\sim\text{Im}\Sigma^{ab}_R(\omega)\propto \mathcal{J}^2\exp\left(-\beta E_g(\frac{q}{2}-1)\right)/\Gamma.
%\end{align}
%This result can also be derived using a semi-classical Boltzmann equation \cite{Kinetics}. This leads to:
%\begin{align}
%\Gamma\sim \mathcal{J}\exp\left(-\beta E_g\frac{(q/2-1)}{2}\right). \label{Gamma}
%\end{align}
%This gives the temperature dependence for $\Gamma$ as a function of $\beta E_g$. For $q=4$, we have $\Gamma\sim \mathcal{J}\exp\left(-\beta E_g/2\right)$. 

\section{Spectral Functions for small $\mathbf{q}$}\label{sec:q4}

We first present the result for small $q$. To be concrete, we choose $q=4$ as an example. We first fix $\mu/\mathcal{J}=0.1$. According to \cite{Xiaoliang}, there would be a first order transition near $\beta \mathcal{J}\approx 35$. In Figure \ref{fig1}, we show the spectral function and Green's functions for $\beta \mathcal{J}=40$ with wormhole geometry (a$_1$) and $\beta \mathcal{J}=30$ with two black holes geometry (b$_1$). 

\begin{figure}[t]
 	\center
 	\includegraphics[width=1\columnwidth]{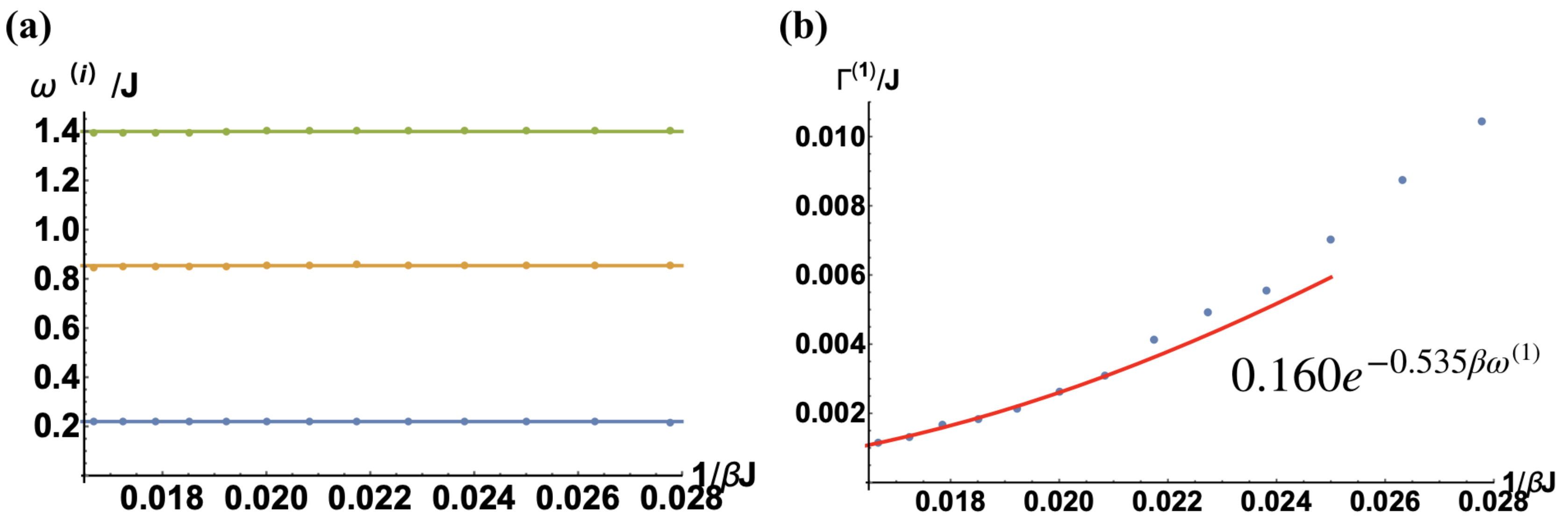}
 	\caption{Numerical results for $q=4$ and $\mu/\mathcal{J}=0.1$. (a). The temperature dependence of the position $\omega^{(i)}$ of the first three peaks. (b). The temperature dependence of the decay rate $\Gamma$ for the lowest mode. The red line represents an exponential fit. } \label{fig2}
 \end{figure}

In the first case $\beta \mathcal{J}=40$, the spectral function is a superposition of narrow peaks with small but finite width, suggesting the existence of well-defined quasi-particles with finite life-time. As a result, the real time Green's function $G^{>,LL}(t)$ (as shown in (Fig. \ref{fig1} a$_2$)) oscillates rapidly for short time but eventually decays in the long-time limit. This is different from the zero-temperature conformal approximation \eqref{GLLT0} without any damping. We could also study the imaginary-time Green's function defined as:
\begin{align}
\hat{G}^{ab}(\tau)=\left<T_\tau\chi^a_i(\tau)\chi^b_i(0)\right>=\begin{pmatrix}
G^{LL}(\tau)&G^{LR}(\tau)\\
G^{RL}(\tau)&G^{RR}(\tau)
\end{pmatrix}.
\end{align}
which is related to the spectral function by:
\begin{align}
\hat{G}(i\omega_n)&=-\int d\omega\frac{\hat{\rho}(\omega)}{i\omega_n-\omega}.
\end{align}
Here $\omega_n=\frac{(2n+1)\pi}{\beta}$ is the Matsubara frequency for fermions. The result for $G^{LL}(\tau)$ and $-iG^{LR}(\tau)$ is shown in Fig. \ref{fig1} (a$_3$), which shows rapid exponential decay, consistent with the presence of an energy gap. We have also checked that our result matches a direct imaginary-time numerics to high accuracy.

For $\beta \mathcal{J}=30$ in the high-temperature phase, the numerical results are very different. The spectral function shown in Figure \ref{fig1} (b$_1$) shows a continuous spectrum similar to that of a single SYK model, except a small splitting near $\omega \sim 0$, due to the relevant coupling $\mu$. As a result, we find $G^{>,LL}(t)$ decays rapidly while $G^{ab}(\tau)$ decays much slower than the previous case.

Since the high temperature phase is more-or-less similar to two decoupled SYK models, we now focus on the low-temperature phase. For fixed $\mu/\mathcal{J}=0.1$, we study the temperature dependence of the energy of the first three peaks $\omega^{(i)},~i=1,2,3$. As shown in Figure \ref{fig2} (a), we find that they are almost independent of temperature, even when the system is close to the transition point. 

To study the life-time of quasi-particles, we define the width $\Gamma^{(i)}$ of $i$-th peak by $\rho^{LL}(\omega^{(i)}-\Gamma^{(i)})=\rho^{LL}(\omega^{(i)})/2$. We focus on the lowest peak with $\omega^{(1)}\equiv E_g$. (According to our estimation \eqref{Gamma}, the higher peaks will have an exponentially smaller width, which is more difficult to study numerically.) %According to \eqref{Gamma}, we expect exponential dependence of temperature at low temperature limit. 
In Fig. \ref{fig2} (b), we show the numerical results for $\Gamma^{(1)}$. Using first seven points, we fit the expression: $\Gamma^{(1)}=a_0 \exp(-c_0/x)$ where $x=1/\beta \mathcal{J}$. We find $c_0=\omega^{(1)}\times 0.535$, close to the analytical prediction. We also check this result holds for $\mu=0.05 \mathcal{J}$. In gravity perspective, this suggests that in the low-temperature phase the system may be identified as weakly interacting bulk fields in global AdS$_2$ background.

\begin{figure}[t]
 	\center
 	\includegraphics[width=1\columnwidth]{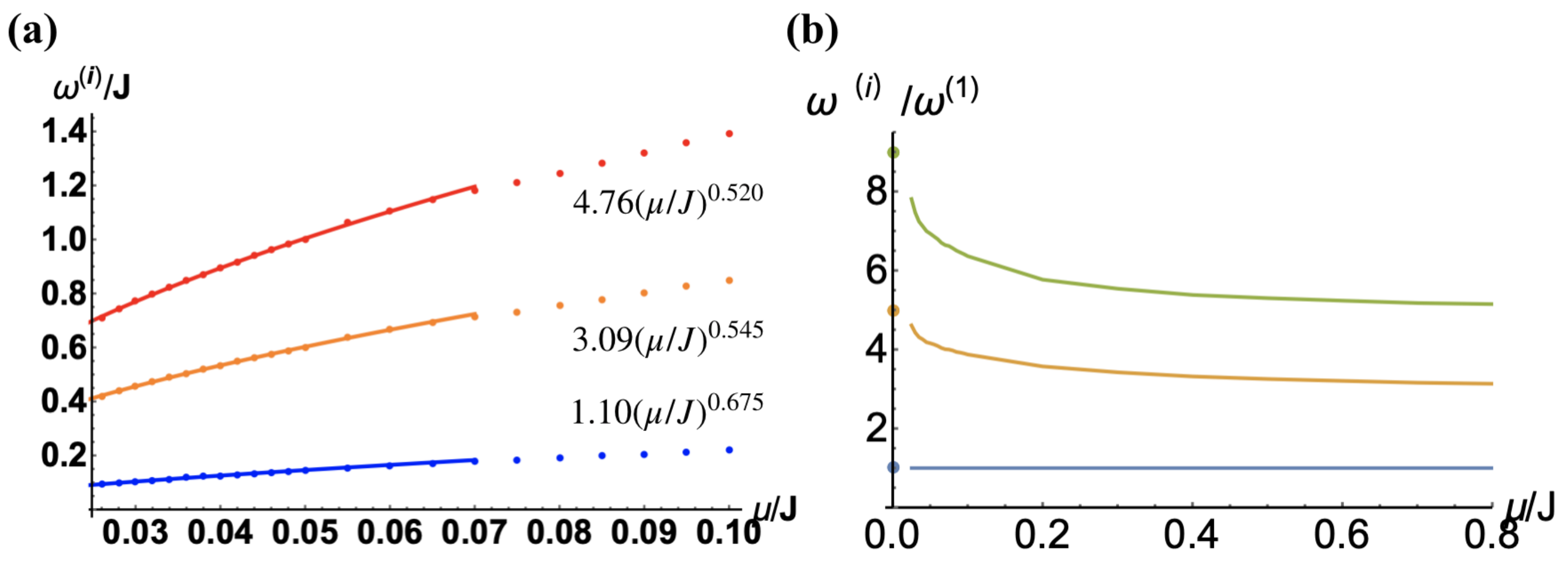}
 	\caption{Numerical results for $q=4$: (a). Positions of the first three peaks $\omega^{(i)}$ as a function of $\mu/J$. The line is a power-law fit. We take $\beta\mathcal{J}=80$ for $\mu/\mathcal{J}<0.05$, while $\beta\mathcal{J}=60$ for other points. (b). energy ratio for the first three peaks as a function of $\mu/J$. The three dots at $\mu/\mathcal{J}=0$ are the theoretical values $1,5,9$.} \label{fig3}
 \end{figure}

 Now we fix the temperature and study the $\mu/\mathcal{J}$ dependence of the energy levels $\omega^{(i)}$. The convergency for iteration is bad deep in the wormhole phase, because the spectral function becomes exponentially narrow. As a result, we only focus on the parameter regions near the phase transition. Fortunately, as we shown in Figure. \ref{fig2} (a), there is almost no temperature dependence for $\omega^{(i)}$. We could then put numerical results with different $\beta \mathcal{J}$ together.

 In Figure \ref{fig3} (a), we plot the positions of first three peaks $\omega^{(i)}$ as a function of $\mu/J$. We take $\beta\mathcal{J}=80$ for $\mu/\mathcal{J}<0.05$, while $\beta\mathcal{J}=60$ for other points. To compare with the analytical result Eq. \eqref{matter}, we fit $\omega^{(i)}=a_ix^{c_i}$, with $x=\mu/\mathcal{J}$ by using the low-temperature data. We find the position of the first mode corresponds to $c_1\sim0.675$, which is close to the analytical result $t'\propto \mu^{2/3}$. However, the other two modes have $c_2\sim 0.545$ and $c_3 \sim 0.520$, which clearly deviate from the low temperature effective theory prediction. In Fig. \ref{fig3} (b), we plot the energy ratios for these first three peaks as a function of $\mu/J$. In the limit of $\mu/\mathcal{J}\rightarrow 0$, the analytical result Eq. \eqref{matter} predict $\omega^{(1)}:\omega^{(2)}:\omega^{(3)}=1:5:9$. Our numerical results are consistent with this prediction, while the second peak approaches $5 \omega^{(1)}$ much quicker than higher modes. On the contrary, the energy ratios approaches $1:3:5$ in the large $\mu/J$ limit where the interaction is a perturbation, which are exactly the energy ratios for free fermion states with 1, 3, 5 particles. 

\section{Spectral Functions at large $\mathbf{q}$}\label{sec:q20}

In the large $q$ limit, Ref. \cite{Xiaoliang} shows that the first order phase transition occurs in a temperature range $\beta \mathcal{J}\propto q\log q$. In this region, the large $q$ solution leads to an exactly periodic Green's function in real time, such that the lifetime of the quasiparticles stays infinite. % studied Different from the small $q$ case where the high temperature phase is close to the decoupled SYK models, it is known that in the $q\rightarrow \infty$ limit the life-time of the system is zero in the wormhole phase even at finite $T$. 
In this section, we numerically study the spectral function for $q=20$, and compare it with the $q=4$ case. 

\begin{figure}[t]
 	\center
 	\includegraphics[width=1\columnwidth]{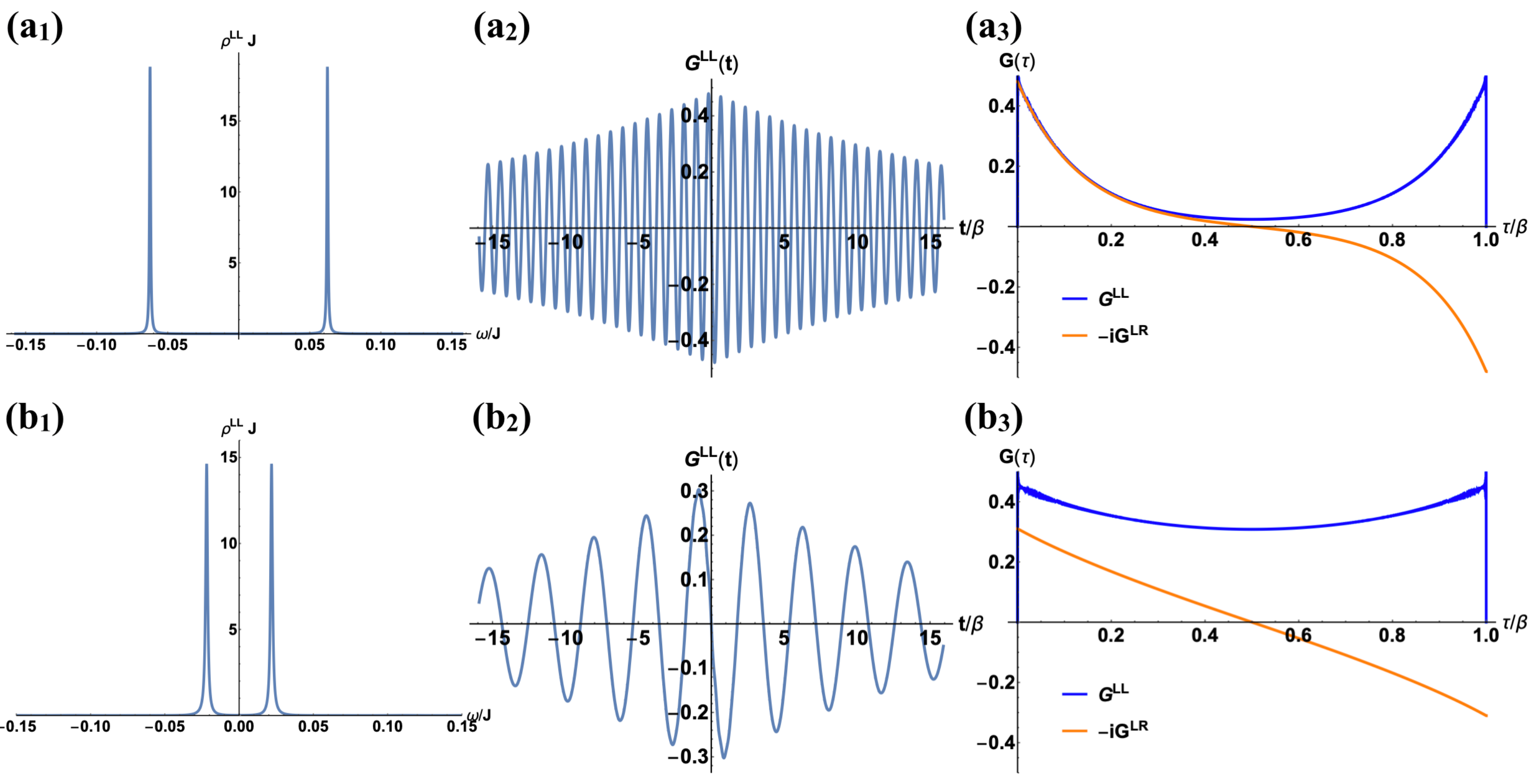}
 	\caption{The spectral function and the Green's function for $q=20$ and $\mu=0.5 \mathcal{J}/q$. (a$_1$-a$_3$). The spectral function $\rho^{LL}(\omega)$, the real-time Green's function $\text{Re}\ G^{>,LL}(t)$ and the imaginary-time Green's function $G^{ab}(\tau)$ for $\beta \mathcal{J}=120$. The system is in the wormhole phase. (b$_1$-b$_3$). The numerical results of the same quantities for $\beta \mathcal{J}=80$. The system is in the two back holes phase.} \label{fig4}
 \end{figure}

 According to Eq. \eqref{GLLT0} and Eq. \eqref{matter}, in the zero-temperature and $\mu/\mathcal{J}\rightarrow 0$ limit, the second peak locates at $\omega^{(2)}=(q+1)\omega^{(1)}$, and its weight is suppressed by $1/q$. As a result, here we could just focus on the lowest mode. 

 As shown in Figure \ref{fig4} (a), the low temperature phase is similar to the $q=4$ case, with an oscillating Green's function in real time and an exponentially decaying Green's function in imaginary time. We take $\mu=0.5 \mathcal{J}/q$ and $\beta \mathcal{J}=120$. The width of the peak is below the numerical resolution limit set by cutoff $\epsilon$ in Eq. \eqref{self}. %of the order of $\epsilon$ cutoff added by hand in Eq. \eqref{self}, which could be approximated by zero within numerical accuracy.

The high temperature phase is very different from the $q=4$ case. The numerical result for $\beta \mathcal{J}=80$ is shown in Figure \ref{fig4} (b). After the first order transition, we only find a jump for $\omega^{(1)}$ and $\Gamma^{(1)}$. There is no qualitative difference from the low-energy phase. We also check that in the high temperature case the width of the peak is larger than the numerical resolution limit $\epsilon$.

\begin{figure}[t]
 	\center
 	\includegraphics[width=1\columnwidth]{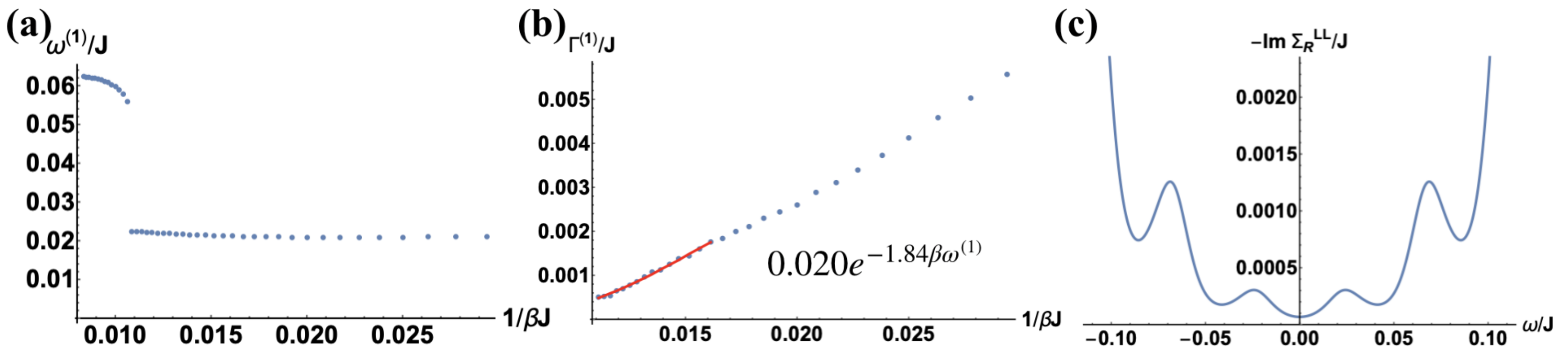}
 	\caption{Numerical results for $\mu=0.5 \mathcal{J}/q$ with $q=20$: (a). The temperature dependence of the energy of the first mode, (b). $\Gamma^{(1)}$ as a function of $\beta \mathcal{J}$, the red line is an exponential fit. (c). The behavior of $\Sigma_R^{LL}(\omega)$ for $\beta \mathcal{J}=80$.} \label{fig5}
 \end{figure}

 The evolution of $\omega^{(1)}$ for both phases and $\Gamma^{(1)}$ for the high temperature phase is shown in Figure \ref{fig5}. Close to the transition point, we find the gap of the low temperature phase drops a little. After the transition, the gap is almost a constant. Since the peak is almost zero-width, the low-energy phase is described by nearly non-interacting bulk fermions in global AdS$_2$ background. In the high temperature phase, the quasi-particle has finite life-time, which is shown in Figure \ref{fig5} (b). However, an exponential fit gives $\Gamma^{(1)}\sim \exp(-1.84 \omega/\beta \mathcal{J})$, different from the analytical approximation $\eqref{Gamma}$. This suggests the interaction is strong and the approximation with only the on-shell processes is not applicable. In other words, the excitation number is not conserved for relevant scattering processes. This can also be seen from directly study the self-energy as shown in (c), where the peak with $\omega \sim 3E_g$ that correspond to an off-shell process is not well-separated from the on-shell peak around $\omega \sim E_g$. 

\section{Conclusion}
\label{sec:conclusion}

In this work, we study the finite temperature spectral function of the coupled SYK model numerically by solving the Schwinger-Dyson equation on the Keldysh contour. 

For small $q=4$, in the low-temperature wormhole phase, we find the system can be described by a weakly interacting Fermion in global AdS$_2$. The spectrum of the fermion consists of sharp quasi-particle peaks with finite lifetime, which is very different from the high temperature phase without quasi-particles. The energy of these modes is almost independent of temperature and their ratio approaches the AdS$_2$ prediction in the small $\mu$ limit. The inverse life time of quasi-particles is consistent with contributions from on-shell scattering processes without the change of particle number, which vanishes exponentially in the low-temperature limit. 

For the large $q=20$, we find that the high temperature phase behaves similar to the low temperature phase, with finite life-time quasi-particles. Across the phase transition, the energy levels experience a jump. In the high temperature phase, the fermions are strongly interacting and the off-shell scattering becomes significant. 

{\bf \noindent Acknowledgement.} We thank Chao-Ming Jian and Shunyu Yao for helpful discussion. After finishing this work, we became aware of an independent work on similar topic\cite{plugge2020}. We would like to thank the authors Stephan Plugge, Etienne Lantagne-Hurtubise and Marcel Franz for sharing their unpublished result with us. This work is supported by the National Science Foundation Grant No. 1720504 and the Simons Foundation. This work is also supported in part by the DOE Office of Science, Office of High Energy Physics, the grant DE-SC0019380.

\bibliography{ref}

\end{document}